# A revisited time domain formulation of boundary integral equations for two-dimensional elastodynamics

Domenico Capuani†

*Università di Ferrara, DA, Via Quartieri 8, 44100 Ferrara, Italy*

**Abstract**

A boundary integral equation (BIE) formulation for 2-D transient elastic wave propagation problems is presented. On the basis of the three-dimensional integral identity, the time-dependent kernels for the two-dimensional boundary integral equation are obtained. A linear time variation of displacements and tractions is assumed over each time step and an implicit time marching scheme is deduced. The formulation is used to obtain an analytical solution for the cylindrical cavity under transient pressure at the boundary surface.

## 1. Introduction

For two-dimensional transient elastodynamic problems, BEM formulations have been proposed in the Laplace domain and in the frequency domain together with a numerical inversion scheme to obtain time domain solutions.

With respect to transform domain formulations, time domain formulations are attractive especially for non-linear problems. As for two-dimensional problems, either two-dimensional [Antes, 1985; Israil and Banerjee, 1990] either three-dimensional [Niwa *et al.*, 1980] time domain elastodynamic kernels have been employed. By using two-dimensional kernels, numerical integration becomes very complicated because the kernels have singularities as well as jumps at the moving wave front. Moreover, the implementation of a numerical time-stepping tecnique requires additional transformations to be carried out in order to eliminate the derivatives of Heaviside functions appearing in the integral equation [Sládek and Sládek, 1992]. When time domain three-dimensional kernels are adopted, difficulties connected with the derivatives of Heaviside function at the moving wave front can be avoided. Niwa et al. [1980] employed numerical techniques to integrate the singular kernels along the third spatial co-ordinate. This procedure leads to a loss of accuracy and stability for small time steps as well as for increasing time intervals of observation.

In this paper, a boundary integral formulation for 2-D transient elastic wave propagation problems is presented. On the basis of the three-dimensional integral identity, taking the causality condition into account and employing the Hadamard method of descent, the convoluted time-dependent kernels for the two-dimensional boundary integral equation are obtained. In this reduction process, a linear time variation of displacements and tractions is assumed over each time step and an implicit time marching scheme is deduced. Among the resulting displacement and traction kernels, the singular ones present the same singularities of the elastostatic case, as the distance between field and source points approaches zero.

† E-mail address: domenico.capuani@unife.it

## 2. The integral equation

In-plane wave motion in a homogeneous, isotropic, linearly elastic solid occupying a two-dimensional regular region *D*, with boundary *C*, is considered. The solid is characterized by Lamé moduli λ and μ, density ρ and propagation velocities $c_1 = [(\lambda+2\mu)/\rho]^{1/2}$ and $c_2 = [\mu/\rho]^{1/2}$ of dilatational and shear waves, respectively. In the absence of body forces, boundary values of displacement and/or traction are prescribed for the solid which is initially at rest.

In a three-dimensional framework, the solid can be seen as an infinite cylindrical body and a Cartesian system of coordinates $\mathbf{x} = (x_1, x_2, x_3)$ is chosen in such a way that the body is infinite along the $x_3$-direction and it occupies the region *D* in the plane $x_3 = 0$. The two-dimensional problem is stated in the $(x_1, x_2)$-plane and the field quantities are independent of $x_3$. At time *t*, the in-plane motion is described by the displacement field $u_i(x_1, x_2, t)$ $(i = 1, 2)$, and the relevant traction $t_{(\mathbf{n})i}$ $(i = 1, 2)$ is given by

$$t_{(\mathbf{n})i}(x_1, x_2, t) = \sigma_{ij} n_j$$

where $\mathbf{n} = (n_1, n_2)$ is the outward unit normal vector and $\sigma_{ij}$ is the stress tensor

$$\sigma_{ij}(x_1, x_2, t) = \rho(c_1^2 - 2c_2^2) u_{m,m} \delta_{ij} + \rho c_2^2 (u_{i,j} + u_{j,i}) \, .$$

Starting from the integral representation for the three-dimensional problem, the in-plane displacement can be given the following integral representation at every point $\mathbf{y} \in$ D (Eringen and Suhubi, 1975):

$$u_k(\mathbf{y}, t) = \int_C \int_{-\infty}^{+\infty} G_{ik}(\mathbf{x} - \mathbf{y}, t) * t_{(\mathbf{n})i}(x_1, x_2, t) dx_3 \, dC(x_1, x_2) - \int_C \int_{-\infty}^{+\infty} S_{(\mathbf{n})ik}(\mathbf{x} - \mathbf{y}, t) * u_i(x_1, x_2, t) dx_3 \, dC(x_1, x_2) \qquad (1)$$

where $G_{ik}$ is the displacement tensor of the three-dimensional Stokes fundamental solution

$$G_{ik}(\mathbf{x}, t) = \frac{1}{4\pi\rho} \left\{ \frac{x_i x_k}{|\mathbf{x}|^3} \left[ \frac{1}{c_1^2} \delta\left(t - \frac{|\mathbf{x}|}{c_1}\right) - \frac{1}{c_2^2} \delta\left(t - \frac{|\mathbf{x}|}{c_2}\right) \right] + \frac{\delta_{ik}}{|\mathbf{x}| c_2^2} \delta\left(t - \frac{|\mathbf{x}|}{c_2}\right) \right.$$

$$\frac{t}{|\mathbf{x}|^2}\left(\frac{3x_i x_k}{|\mathbf{x}|^3}-\frac{\delta_{ik}}{|\mathbf{x}|}\right)\left[H\left(t-\frac{|\mathbf{x}|}{c_1}\right)-H\left(t-\frac{|\mathbf{x}|}{c_2}\right)\right]\Bigg\}$$

$S_{(\mathbf{n})ik}$ is the relevant traction tensor

$$S_{(\mathbf{n})ik}(\mathbf{x},t)=\rho(c_1^2-2c_2^2)G_{mk,m}n_i+\rho c_2^2\big(G_{ik,j}+G_{jk,i}\big)n_j$$

and the symbol * denotes the convolution on the time variable, i.e for any given function $f(x, t)$, $g(x, t)$:

$$f*g=\int_0^t f(x,\tau)g(x,t-\tau)\,d\tau \tag{2}$$

## 2. Time-convoluted kernels for the two-dimensional problem

In order to perform the time convolution in Eq. (1), the time interval $(0, t_N)$ is divided into $N$ time steps $(t_{n-1}, t_n)$ for $n = 1, 2, ..., N$. Within each time step, the following linear representation for displacements and tractions is adopted:

$$u_i(x_1,x_2,t)=u_i(x_1,x_2,t_{n-1})\psi_1(\tau)+u_i(x_1,x_2,t_n)\psi_2(\tau) \tag{3a}$$

$$t_{(\mathbf{n})i}(x_1,x_2,t)=t_{(\mathbf{n})i}(x_1,x_2,t_{n-1})\psi_1(\tau)+t_{(\mathbf{n})i}(x_1,x_2,t_n)\psi_2(\tau) \tag{3b}$$

where $\tau\in[t_{n-1}, t_n]$,

$$\psi_1(\tau)=\frac{t_n-\tau}{\Delta t_n},\qquad \psi_2(\tau)=\frac{\tau-t_{n-1}}{\Delta t_n} \tag{4}$$

and $\Delta t_n=t_n-t_{n-1}$. Substituting Eqs. (3), and observing that

$$u_i(x_1,x_2,t_N-\tau)=u_i(x_1,x_2,t_N-t_n)\psi_2(\tau)+u_i(x_1,x_2,t_N-t_{n-1})\psi_1(\tau) \tag{5a}$$

$$t_{(\mathbf{n})i}(x_1,x_2,t_N-\tau)=t_{(\mathbf{n})i}(x_1,x_2,t_N-t_n)\psi_2(\tau)+t_{(\mathbf{n})i}(x_1,x_2,t_N-t_{n-1})\psi_1(\tau) \tag{5b}$$

the following time-marching scheme can be obtained from Eq. (1):

$$u_k(\mathbf{y},t)-\int_C\big[U_{ik}^{11}t_{(\mathbf{n})i}(x_1,x_2,t_N)-T_{(\mathbf{n})ik}^{11}u_i(x_1,x_2,t_N)\big]dC=\sum_{n=2}^{N}\int_C\big[U_{ik}^{1n}t_{(\mathbf{n})i}(x_1,x_2,t_N-t_{n-1})-T_{(\mathbf{n})ik}^{1n}u_i(x_1,x_2,t_N-t_{n-1})\big]dC \tag{6}$$

$$+\sum_{n=1}^{N}\int_C \left[U_{ik}^{2n} t_{(\mathbf{n})i}(x_1,x_2,t_N-t_n) - T_{(\mathbf{n})ik}^{1n} u_i(x_1,x_2,t_N-t_n)\right]dC$$

where, according to the symmetry with respect to the plane $x_3 = 0$, the kernels associated with the $n$-th time step are defined as ($\alpha = 1, 2$):

$$U_{ik}^{\alpha n} = 2\int_0^{+\infty} dx_3 \int_{t_{n-1}}^{t_n} G_{ik}(\mathbf{x}-\mathbf{y},\tau)\psi_\alpha(\tau)d\tau \tag{7a}$$

$$T_{(\mathbf{n})ik}^{\alpha n} = 2\int_0^{+\infty} dx_3 \int_{t_{n-1}}^{t_n} S_{(\mathbf{n})ik}(\mathbf{x}-\mathbf{y},\tau)\psi_\alpha(\tau)d\tau \tag{7b}$$

By performing the time and space integrations in the R.H.S. of Eq. (7a), taking the causality principle into account, the analytical expression for the time-convoluted displacement kernel is obtained in the form:

$$U_{ik}^{\alpha n} = \frac{1}{2\pi\rho}\left\{\frac{1}{c_1^2}A_{ik}^{\alpha n} + \frac{1}{c_1^2}B_{ik}^{\alpha n} + \left[\frac{1}{c_1^2}C_{ik}^{\alpha n}(c_1) - \frac{1}{c_2^2}C_{ik}^{\alpha n}(c_2)\right]\right\} \tag{8}$$

where

$$A_{ik}^{\alpha n} = (-1)^\alpha \frac{r_i r_k}{r}\left\{-\frac{1}{r}\frac{t_\alpha}{\Delta t_n}[f_1(c_1,t_n) - f_1(c_1,t_{n-1})] + \frac{1}{c_1\Delta t_n}[f_2(c_1,t_n) - f_2(c_1,t_{n-1})]\right\} \tag{9a}$$

$$B_{ik}^{\alpha n} = (-1)^\alpha \left\{\frac{r_i r_k}{r^2}\frac{t_\alpha}{\Delta t_n}[f_1(c_2,t_n) - f_1(c_2,t_{n-1})] - \frac{r_i r_k}{r}\frac{1}{c_2\Delta t_n}[f_2(c_2,t_n) - f_2(c_2,t_{n-1})] + \frac{\delta_{ik}}{\Delta t_n}[t_n f_1(c_2,t_n) - t_{n-1}f_1(c_2,t_{n-1})] + \frac{\delta_{ik}}{\Delta t_n}t_\alpha[f_3(c_2,t_n) - f_3(c_2,t_{n-1})]\right\} \tag{9b}$$

$$\begin{aligned}C_{ik}^{\alpha n}(c) = (-1)^\alpha \Bigg\{&\frac{r_i r_k}{r^2}\frac{t_\alpha}{\Delta t_n}[f_1(c,t_n) - f_1(c,t_{n-1})] - \frac{r_i r_k}{r}\frac{1}{c\Delta t_n}[f_2(c,t_n) - f_2(c,t_{n-1})]\\ &+ \left(\frac{r_i r_k}{r^2} + \delta_{ik}\right)\frac{1}{3\Delta t_n}[t_n f_1(c,t_n) - t_{n-1}f_1(c,t_{n-1})]\\ &+ \delta_{ik}\frac{t_\alpha}{2\Delta t_n}[f_3(c,t_n) - f_3(c,t_{n-1})] - \left(2\frac{r_i r_k}{r^2} - \delta_{ik}\right)\cdot\\ &\cdot\frac{1}{r^2}\frac{c^2}{6\Delta t_n}[t_n^2(3t_\alpha - 2t_n)f_1(c,t_n) - t_{n-1}^2(3t_\alpha - 2t_{n-1})f_1(c,t_{n-1})]\Bigg\}\end{aligned} \tag{9c}$$

with $t_{\alpha=1}=t_n$, $t_{\alpha=2}=t_{n-1}$, $r_i=x_i-y_i$, $r=(r_1^2+r_2^2)^{1/2}$,

$$f_1(c,t)=H\left(t-\frac{r}{c}\right)\sqrt{1-\frac{r^2}{c^2t^2}} \tag{10a}$$

$$f_2(c,t)=H\left(t-\frac{r}{c}\right)\tan^{-1}\left(\frac{ct}{r}\sqrt{1-\frac{r^2}{c^2t^2}}\right) \tag{10b}$$

$$f_3(c,t)=H\left(t-\frac{r}{c}\right)\log\left\{r\Big/\left[ct\left(1+\sqrt{1-\frac{r^2}{c^2t^2}}\right)\right]\right\} \tag{10c}$$

and $H(\cdot)$ is the Heaviside function.

Analogously, from Eq. (7b) the following analytical expression is determined for the time-dependent traction kernel:

$$T_{(\mathbf{n})ik}^{\alpha n}=\frac{1}{2\pi}\{\Gamma_{ik}^{\alpha n}+\Delta_{ik}^{\alpha n}+[E_{ik}^{\alpha n}(c_1)-E_{ik}^{\alpha n}(c_2)]+H_{ik}^{\alpha n}+\Theta_{ik}^{\alpha n}\} \tag{11}$$

where

$$\begin{aligned}\Gamma_{ik}^{\alpha n}=(-1)^{\alpha}\Bigg\{&\left[\frac{r_kn_i}{r^2}-2\frac{c_2^2}{c_1^2}\left(\frac{r_mn_m\delta_{ik}+r_in_k+2r_kn_i}{r^2}-\frac{r_mn_mr_ir_k}{r^4}\right)\right]\cdot\\
&\cdot\frac{t_\alpha}{\Delta t_n}[f_1(c_1,t_n)-f_1(c_1,t_{n-1})]\\
&-6\frac{c_2^2}{c_1^2}\frac{r_mn_mr_ir_k}{r^2}\frac{1}{c_1^2\Delta t_n}\left[\frac{f_1(c_1,t_n)}{t_n}-\frac{f_1(c_1,t_{n-1})}{t_{n-1}}\right]\\
&+4\frac{c_2^2}{c_1^2}\frac{r_mn_mr_ir_k}{r^2}\frac{1}{c_1^2}\frac{t_\alpha}{\Delta t_n}\left[\frac{f_1(c_1,t_n)}{t_n^2}-\frac{f_1(c_1,t_{n-1})}{t_{n-1}^2}\right]\\
&-\left[\frac{r_kn_i}{r}-2\frac{c_2^2}{c_1^2}\left(\frac{r_mn_m\delta_{ik}+r_in_k+2r_kn_i}{r}-3\frac{r_mn_mr_ir_k}{r^3}\right)\right]\times\\
&\times\frac{f_2(c_1,t_n)-f_2(c_1,t_{n-1})}{c_1\Delta t_n}\end{aligned} \tag{12a}$$

$$\begin{aligned}\Delta_{ik}^{\alpha n}=(-1)^{\alpha}\Bigg\{&\left(\frac{3r_mn_m\delta_{ik}+3r_in_k+2r_kn_i}{r^2}-8\frac{r_mn_mr_ir_k}{r^4}\right)\cdot\\
&\cdot\frac{t_\alpha}{\Delta t_n}[f_1(c_2,t_n)-f_1(c_2,t_{n-1})]\\
&+6\frac{r_mn_mr_ir_k}{r^2}\frac{1}{c_2^2\Delta t_n}\left[\frac{f_1(c_2,t_n)}{t_n}-\frac{f_1(c_2,t_{n-1})}{t_{n-1}}\right]\\
&-4\frac{r_mn_mr_ir_k}{r^2}\frac{1}{c_2^2}\frac{t_\alpha}{\Delta t_n}\left[\frac{f_1(c_2,t_n)}{t_n^2}-\frac{f_1(c_2,t_{n-1})}{t_{n-1}^2}\right]\end{aligned}$$

$$-\left(\frac{3r_m n_m \delta_{ik}+3r_i n_k + 2r_k n_i}{r} - 6\frac{r_m n_m r_i r_k}{r^3}\right)\frac{f_2(c_2,t_n)-f_2(c_2,t_{n-1})}{c_2 \Delta t_n}\Bigg\} \quad (12b)$$

$$\mathrm{E}_{ik}^{\alpha n}(c) = (-1)^\alpha \Bigg\{\left(\frac{r_m n_m \delta_{ik}+r_i n_k + r_k n_i}{r^2} - 3\frac{r_m n_m r_i r_k}{r^4}\right)\cdot$$
$$\cdot\frac{2c_2^2}{c^2}\frac{t_\alpha}{\Delta t_n}[f_1(c,t_n)-f_1(c,t_{n-1})]$$
$$+3\frac{r_m n_m r_i r_k}{r^2}\frac{c_2^2}{c^4 \Delta t_n}\left[\frac{f_1(c,t_n)}{t_n}-\frac{f_1(c,t_{n-1})}{t_{n-1}}\right]$$
$$-2\frac{r_m n_m r_i r_k}{r^2}\frac{c_2^2}{c^4}\frac{t_\alpha}{\Delta t_n}\left[\frac{f_1(c,t_n)}{t_n^2}-\frac{f_1(c,t_{n-1})}{t_{n-1}^2}\right]$$
$$-\left(2\frac{r_m n_m \delta_{ik}+r_i n_k + r_k n_i}{r} - 5\frac{r_m n_m r_i r_k}{r^3}\right)c_2^2\frac{f_2(c,t_n)-f_2(c,t_{n-1})}{c^3\Delta t_n}$$
$$+\left(\frac{r_m n_m \delta_{ik}+r_i n_k + r_k n_i}{r^2} - 4\frac{r_m n_m r_i r_k}{r^4}\right)\frac{2}{3}c_2^2\frac{t_n f_1(c,t_n)-t_{n-1}f_1(c,t_{n-1})}{c^2\Delta t_n}$$
$$-\left(\frac{r_m n_m \delta_{ik}+r_i n_k + r_k n_i}{r^4} - 4\frac{r_m n_m r_i r_k}{r^6}\right)\cdot$$
$$\cdot\frac{2}{3}c_2^2\frac{(3t_\alpha-2t_n)t_n^2 f_1(c,t_n)-(3t_\alpha-2t_{n-1})t_{n-1}^2 f_1(c,t_{n-1})}{\Delta t_n}\Bigg\} \quad (12c)$$

$$\mathrm{H}_{ik}^{\alpha n} = (-1)^\alpha \Bigg\{\frac{c_2^2}{c_1^2}\frac{r_m n_m r_i r_k}{r^2}\frac{1}{c_1^2 \Delta t_n}\left[\frac{f_1(c_1,t_n)}{t_n}-\frac{f_1(c_1,t_{n-1})}{t_{n-1}}\right]$$
$$-\left(\frac{(2c_2^2-c_1^2)}{c_1^2}\frac{r_k n_i}{r}-\frac{c_2^2}{c_1^2}\frac{r_m n_m r_i r_k}{r^3}\right)\frac{f_2(c_1,t_n)-f_2(c_1,t_{n-1})}{c_1\Delta t_n}\Bigg\} \quad (12d)$$

$$\Theta_{ik}^{\alpha n} = (-1)^\alpha \Bigg\{-\frac{r_m n_m r_i r_k}{r^2}\frac{1}{c_2^2 \Delta t_n}\left[\frac{f_1(c_2,t_n)}{t_n}-\frac{f_1(c_2,t_{n-1})}{t_{n-1}}\right]$$
$$+\left(\frac{r_m n_m \delta_{ik}+r_i n_k}{r}-\frac{r_m n_m r_i r_k}{r^3}\right)\frac{f_2(c_2,t_n)-f_2(c_2,t_{n-1})}{c_2\Delta t_n}\Bigg\} \quad (12e)$$

It can be seen that kernels (8, 11) present singularities as $r \to 0$ only in the first time-step, i.e. for $n = 1$, whereas they are bounded in the successive time steps ($n > 1$). Moreover, the singularities are the same of the elastostatic case, namely of type $\log(r)$ and $(1/r)$ for the displacement and the traction kernel respectively.

The boundary integral equation (6) gives the instant value of the displacement field in the interior points of the body, once the time-histories of boundary values of displacement and traction are known.

When the source point $\mathbf{y}$ is taken to the boundary $C$, Eq. (6) is re-written as

$$C_{ik}(\mathbf{y})u_k(\mathbf{y},t) + PV\int_C T_{(\mathbf{n})ik}^{11}u_i(x_1,x_2,t_N)dC - \int_C U_{ik}^{11}t_{(\mathbf{n})i}(x_1,x_2,t_N)dC =$$
$$\sum_{n=2}^{N}\int_C \left[U_{ik}^{1n}t_{(\mathbf{n})i}(x_1,x_2,t_N-t_{n-1}) - T_{(\mathbf{n})ik}^{1n}u_i(x_1,x_2,t_N-t_{n-1})\right]dC \quad (13)$$

$$+\sum_{n=1}^{N}\int_{C}\left[U_{ik}^{2n}t_{(\mathbf{n})i}(x_1,x_2,t_N-t_n)-T_{(\mathbf{n})ik}^{1n}u_i(x_1,x_2,t_N-t_n)\right]dC$$

where $PV$ stands for principal value of the integral and

$$C_{ik}(\mathbf{y})=\lim_{\varepsilon\to 0}\int_{\Gamma_\varepsilon}T_{(\mathbf{n})ik}^{11}dC \tag{14}$$

In Eq. (14), $\Gamma_\varepsilon$ is the intersection of the circle of radius ε, centered at **y**, with the region occupied by the body and, at a smooth point of the boundary, it turns out to be $C_{ik}=\delta_{ij}/2$.

Equation (13) can be employed in a time-domain boundary element formulation to solve in-plane boundary value problems of transient wave propagation.

## 3. The cylindrical cavity

A circular cylindrical cavity, of radius $R$ and of infinite length, in an infinite elastic medium is considered. At the time $t$ = 0 the solid is supposed to be in the unstressed state and at rest. At this time, a uniform pressure with arbitrary time-variation $p(t)$ is applied at the surface of the cavity. The corresponding radial displacement $u(t)$ of cavity surface at any time $t$, can be obtained by integrating Eq. (13) analytically. To this purpose, a system of polar coordinates ($R$, θ) with the origin at the centre of the cavity is introduced, so as to write the distance $r$, between field and source point, the outward unit normal $n_i$ to the cavity surface, the displacement $u_i$ and the traction $t_{(\mathbf{n})i}$ as

$$r=2R\sin\frac{\theta}{2},\quad 0\le\theta\le\pi \tag{15a}$$

$$r_1=-R(1-\cos\theta),\qquad r_2=R\sin\theta \tag{15b}$$

$$n_1=-\cos\theta,\qquad n_2=-\sin\theta \tag{15c}$$

$$u_1=u\cos\theta,\qquad u_2=u\sin\theta \tag{15d}$$

$$t_{(\mathbf{n})1}=-pn_1=p\cos\theta,\qquad t_{(\mathbf{n})2}=-pn_2=p\sin\theta \tag{15e}$$

Accordingly, Eq. (13) becomes

$$\left(\frac{1}{2}+a^{11}\right)u(t_N)=b^{11}p(t_N)+R^{N-1} \tag{16}$$

where $R^{N-1}$ collects the contributions of all the previous N-1 time steps

$$R^{N-1}=\sum_{n=2}^{N}[b^{1n}p(t_N-t_{n-1})-a^{1n}u(t_N-t_{n-1})]+$$

$$+\sum_{n=1}^{N}[b^{2n}p(t_N-t_n)-a^{2n}u(t_N-t_n)] \quad (17)$$

and coefficients $a^{\alpha n}$, $b^{\alpha n}$ are defined as

$$a^{\alpha n}=2\int_0^{\pi}(T_{11}^{\alpha n}(R,\theta)\cos\theta+T_{21}^{\alpha n}(R,\theta)\sin\theta)Rd\theta \quad (18a)$$

$$b^{\alpha n}=2\int_0^{\pi}(U_{11}^{\alpha n}(R,\theta)\cos\theta+U_{21}^{\alpha n}(R,\theta)\sin\theta)Rd\theta \quad (18b)$$

Taking Eq. (8) into account, Eq. (18b) can be written as

$$b^{\alpha n}=\frac{2}{2\pi\rho}\left\{\frac{1}{c_1^2}\left(A^{\alpha n}(c_1,t_n)-A^{\alpha n}(c_1,t_{n-1})\right)\right.$$

$$+\frac{1}{c_2^2}\left(B^{\alpha n}(c_2,t_n)-B^{\alpha n}(c_2,t_{n-1})\right)$$

$$\left.+\left[\frac{1}{c_1^2}\left(C^{\alpha n}(c_1,t_n)-C^{\alpha n}(c_1,t_{n-1})\right)-\frac{1}{c_2^2}\left(C^{\alpha n}(c_2,t_n)-C^{\alpha n}(c_2,t_{n-1})\right)\right]\right\} \quad (19)$$

where

$$A^{\alpha n}=\int_0^{\pi}(A_{11}^{\alpha n}(R,\theta)\cos\theta+A_{21}^{\alpha n}(R,\theta)\sin\theta)Rd\theta \quad (20a)$$

$$B^{\alpha n}=\int_0^{\pi}(B_{11}^{\alpha n}(R,\theta)\cos\theta+B_{21}^{\alpha n}(R,\theta)\sin\theta)Rd\theta \quad (20b)$$

$$C^{\alpha n}=\int_0^{\pi}(C_{11}^{\alpha n}(R,\theta)\cos\theta+C_{21}^{\alpha n}(R,\theta)\sin\theta)Rd\theta \quad (20c)$$

Integrating Eqs. (20) gives

$$A^{\alpha n}(c,t)=(-1)^{\alpha}\cdot 2\left\{\frac{t_\alpha}{\Delta t_n}\frac{1}{3k^2}[(2k^2-1)E(\varphi/2,k)-(k^2-1)F(\varphi/2,k)]\right.$$

$$\left.-\frac{R}{c\Delta t_n}\frac{1}{9k^3}[-2(2+5k^2)E(\varphi/2,k)+4(1+2k^2-3k^4)F(\varphi/2,k)]\right\} \quad (21a)$$

$$B^{\alpha n}(c,t)=-A^{\alpha n}(c,t)$$

$$+(-1)^{\alpha}\cdot 2\left\{\frac{t}{\Delta t_n}\frac{1}{3k^2}\left[(2-k^2)E\left(\frac{\varphi}{2},k\right)+2(k^2-1)F(\varphi/2,k)\right]\right.$$

$$-\frac{t_\alpha}{\Delta t_n}\frac{1}{k^2}[E(\varphi/2,k)+(k^2-1)F(\varphi/2,k)]\Big\} \quad (21b)$$

$$C^{\alpha n}(c,t) = -A^{\alpha n}(c,t) \\ +(-1)^\alpha\cdot 2\left\{\frac{t}{3\Delta t_n}\frac{1}{k^2}\left[(1-k^2)E\left(\frac{\varphi}{2},k\right)+(k^2-1)F(\varphi/2,k)\right]\right. \\ -\frac{t_\alpha}{2\Delta t_n}\frac{1}{k^2}[E(\varphi/2,k)+(k^2-1)F(\varphi/2,k)] \\ \left.-\frac{c^2t^2}{4R^2}\frac{(3t_\alpha-2t)}{6\Delta t_n}[E(\varphi/2,k)+(k^2-1)F(\varphi/2,k)]\right\} \quad (21c)$$

where

$$F(\varphi/2,k) = \int_0^{\varphi/2}\frac{d\beta}{\sqrt{1-k^2\sin^2\beta}}\,, \qquad E(\varphi/2,k) = \int_0^{\varphi/2}\sqrt{1-k^2\sin^2\beta}\,d\beta \quad (22)$$

are normal elliptic integrals of the first kind and of the second kind (Erdelyi, 1954)), respectively, with modulus $k$ and amplitude $\varphi/2$ defined as

$$k(c,t) = \frac{2R}{ct} \quad (23a)$$

$$\varphi(c,t) = \begin{cases} 2\arcsin\frac{1}{k} & \text{if } k>1 \\ \pi & \text{if } k\le 1 \end{cases} \quad (23b)$$

Analogously, taking Eq. (11) into account, Eq. (18a) can be written as

$$a^{\alpha n} = \frac{2}{2\pi}\left\{\left(\Gamma^{\alpha n}(c_1,t_n)-\Gamma^{\alpha n}(c_1,t_{n-1})\right)+\left(\Delta^{\alpha n}(c_2,t_n)-\Delta^{\alpha n}(c_2,t_{n-1})\right)\right. \\ \left[\left(E^{\alpha n}(c_1,t_n)-E^{\alpha n}(c_1,t_{n-1})\right)-\left(E^{\alpha n}(c_2,t_n)-E^{\alpha n}(c_2,t_{n-1})\right)\right] \\ \left.+\left(\mathrm{H}^{\alpha n}(c_1,t_n)-\mathrm{H}^{\alpha n}(c_1,t_{n-1})\right)+\left(\Theta^{\alpha n}(c_2,t_n)-\Theta^{\alpha n}(c_2,t_{n-1})\right)\right\} \quad (24)$$

where

$$\Gamma^{\alpha n} = \int_0^\pi(\Gamma_{11}^{\alpha n}(R,\theta)\cos\theta+\Gamma_{21}^{\alpha n}(R,\theta)\sin\theta)Rd\theta \quad (25a)$$

$$\Delta^{\alpha n} = \int_0^\pi(\Delta_{11}^{\alpha n}(R,\theta)\cos\theta+\Delta_{21}^{\alpha n}(R,\theta)\sin\theta)Rd\theta \quad (25b)$$

$$E^{\alpha n} = \int_0^\pi(E_{11}^{\alpha n}(R,\theta)\cos\theta+E_{21}^{\alpha n}(R,\theta)\sin\theta)Rd\theta \quad (25c)$$

$$H^{\alpha n} = \int_0^\pi(H_{11}^{\alpha n}(R,\theta)\cos\theta+H_{21}^{\alpha n}(R,\theta)\sin\theta)Rd\theta \quad (25d)$$

$$\Theta^{\alpha n} = \int_0^{\pi} (\Theta_{11}^{\alpha n}(R,\theta)\cos\theta + \Theta_{21}^{\alpha n}(R,\theta)\sin\theta) R d\theta \tag{25e}$$

Integrating Eqs. (25) gives

$$\begin{aligned}
\Gamma^{\alpha n}(c,t) = (-1)^{\alpha} 2 \Bigg\{ & \frac{1}{2R}\frac{t_\alpha}{\Delta t_n}\Big[E(\frac{\varphi}{2},k^2) \\
& -2\frac{c_2^2}{c_1^2}\frac{1}{k^2}\Big((2k^2-1)E\left(\frac{\varphi}{2},k^2\right) + (1-k^2)F(\frac{\varphi}{2},k^2)\Big)\Big] \\
& -6\frac{c_2^2}{c_1^2}\frac{1}{c_1^2\Delta t_n t}\frac{2R}{15k^4}\Big[(8k^4-3k^2-2)E\left(\frac{\varphi}{2},k^2\right) - 2(2k^4-k^2-1)F\left(\frac{\varphi}{2},k^2\right)\Big] \\
& +4\frac{c_2^2}{c_1^4}\frac{t_\alpha}{\Delta t_n t^2}\frac{2R}{15k^4}\Big[(8k^4-3k^2-2)E\left(\frac{\varphi}{2},k^2\right) - 2(2k^4-k^2-1)F\left(\frac{\varphi}{2},k^2\right)\Big] \\
& -\frac{1}{c_1\Delta t_n}\Big[\frac{1}{k}\Big(-E\left(\frac{\varphi}{2},k^2\right) - (m-1)F\left(\frac{\varphi}{2},k^2\right)\Big) \\
& -2\frac{c_2^2}{c_1^2}\frac{1}{9k^2}\Big(-(2+5k^2)E\left(\frac{\varphi}{2},k^2\right) + 2(1+2k^2-3k^4)F\left(\frac{\varphi}{2},k^2\right)\Big)\Big]\Bigg\}
\end{aligned} \tag{26a}$$

$$\begin{aligned}
\Delta^{\alpha n}(c,t) = (-1)^{\alpha} 2 \Bigg\{ & \frac{t_\alpha}{\Delta t_n}\Big[-(1+k^2)E\left(\frac{\varphi}{2},k^2\right) + (1-k^2)F\left(\frac{\varphi}{2},k^2\right)\Big] \\
& +6\frac{1}{c_2^2\Delta t_n t}\frac{2R}{15k^4}\Big[(8k^4-3k^2-2)E\left(\frac{\varphi}{2},k^2\right) - 2(2k^4-k^2-1)F\left(\frac{\varphi}{2},k^2\right)\Big] \\
& -4\frac{1}{c_2^2}\frac{t_\alpha}{\Delta t_n t^2}\frac{2R}{15k^4}\Big[(8k^4-3k^2-2)E\left(\frac{\varphi}{2},k^2\right) - 2(2k^4-k^2-1)F\left(\frac{\varphi}{2},k^2\right)\Big] \\
& -\frac{1}{c_2\Delta t_n}\frac{2}{3k^3}\Big[(k^2-2)E\left(\frac{\varphi}{2},k^2\right) - 2(k^2-1)F\left(\frac{\varphi}{2},k^2\right)\Big]\Bigg\}
\end{aligned} \tag{26b}$$

$$\begin{aligned}
E^{\alpha n}(c,t) = (-1)^{\alpha} 2 \Bigg\{ & \frac{2c_2^2}{c^2}\frac{t_\alpha}{\Delta t_n}\frac{1}{6Rk^2}\Big[-(1+k^2)E\left(\frac{\varphi}{2},k^2\right) \\
& +(1-k^2)F\left(\frac{\varphi}{2},k^2\right)\Big] \\
& +3\frac{c_2^2}{c^4\Delta t_n t}\frac{2R}{15k^4}\Big[(8k^4-3k^2-2)E\left(\frac{\varphi}{2},k^2\right) - 2(2k^4-k^2-1)F\left(\frac{\varphi}{2},k^2\right)\Big] \\
& -2\frac{c_2^2}{c^4}\frac{t_\alpha}{\Delta t_n t^2}\frac{2R}{15k^4}\Big[(8k^4-3k^2-2)E\left(\frac{\varphi}{2},k^2\right) - 2(2k^4-k^2-1)F\left(\frac{\varphi}{2},k^2\right)\Big] \\
& -\frac{c_2^2}{c^3\Delta t_n}\frac{1}{3k^3}\Big[(k^2-2)E\left(\frac{\varphi}{2},k^2\right) - 2(k^2-1)F\left(\frac{\varphi}{2},k^2\right)\Big] \\
& -\frac{2}{3}\frac{c_2^2 t}{c^2\Delta t_n}\frac{1}{2R}E\left(\frac{\varphi}{2},k^2\right) \\
& -\frac{2}{3}c_2^2\frac{(3t_\alpha-2t)t^2}{\Delta t_n}\frac{1}{8R^3}\Big[E\left(\frac{\varphi}{2},k^2\right) + (k^2-1)F\left(\frac{\varphi}{2},k^2\right)\Big]\Bigg\}
\end{aligned} \tag{26c}$$

$$H^{\alpha n}(c,t) = (-1)^{\alpha} 2 \left\{ \frac{c_2^2}{c_1^2} \frac{1}{c_1^2 \Delta t_n} \frac{1}{t} \frac{2R}{15k^4} \left[ (8k^4 - 3k^2 - 2) E\left(\frac{\varphi}{2}, k^2\right) \right. \right.$$
$$\left. -2(2k^4 - k^2 - 1) F\left(\frac{\varphi}{2}, k^2\right) \right]$$
$$- \frac{1}{c_1 \Delta t_n} \frac{(2c_2^2 - c_1^2)}{c_1^2} \frac{1}{k} \left[ -E\left(\frac{\varphi}{2}, k^2\right) - (m-1) F\left(\frac{\varphi}{2}, k^2\right) \right]$$
$$\left. - \frac{1}{c_1 \Delta t_n} \frac{c_2^2}{c_1^2} \frac{1}{9k^3} \left( -(2 + 5k^2) E\left(\frac{\varphi}{2}, k^2\right) + 2(1 + 2k^2 - 3k^4) F\left(\frac{\varphi}{2}, k^2\right) \right) \right\} \quad (26d)$$

$$\Theta^{\alpha n}(c,t) = (-1)^{\alpha} 2 \left\{ - \frac{1}{c_2^2 \Delta t_n} \frac{1}{t} \frac{2R}{15k^4} \left[ (8k^4 - 3k^2 - 2) E\left(\frac{\varphi}{2}, k^2\right) \right. \right.$$
$$\left. -2(2k^4 - k^2 - 1) F\left(\frac{\varphi}{2}, k^2\right) \right]$$
$$\left. + \frac{1}{c_2 \Delta t_n} \frac{1}{3k^3} \left( (k^2 - 2) E\left(\frac{\varphi}{2}, k^2\right) - 2(k^2 - 1) F\left(\frac{\varphi}{2}, k^2\right) \right) \right\} \quad (26e)$$

Equation (16) was used to solve the case of a step pressure pulse $p(t) = p \cdot H(t)$ applied at the surface of the cavity (Figure 1). The stability of the solution was checked for different values of the time-step. The results obtained for $\lambda/\mu = 1$ adopting a time-step $\Delta t = 0.5R/c_1$ are compared with the solution given by Selberg (1952) in Figure 2, where the normal stress $\sigma_T$ in the tangential direction on the cavity contour versus the non-dimensional time parameter $(t \cdot c_1/R)$ is reported ($R = a$ in the Figure).

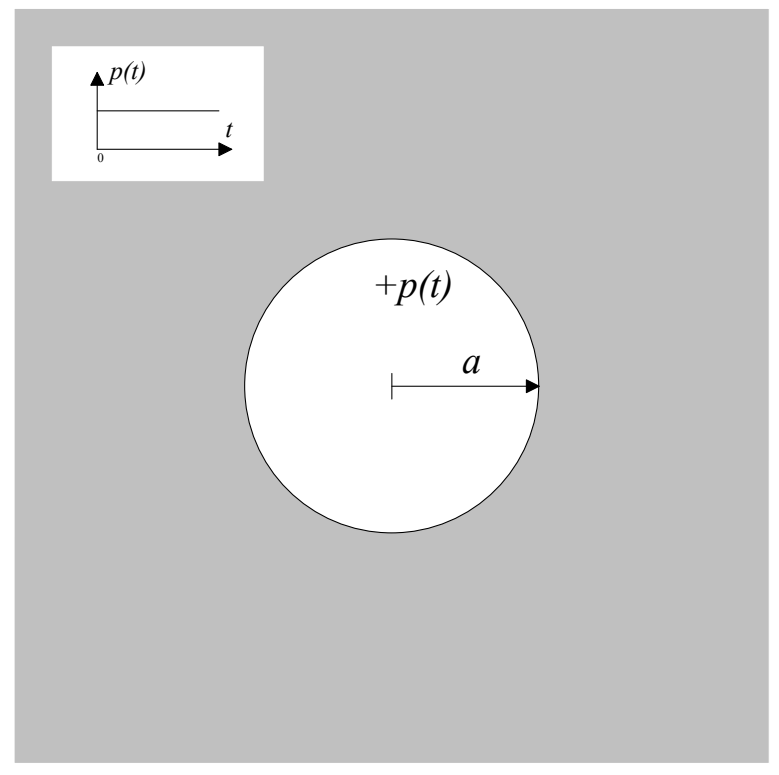

Figure 1: Cylindrical cavity

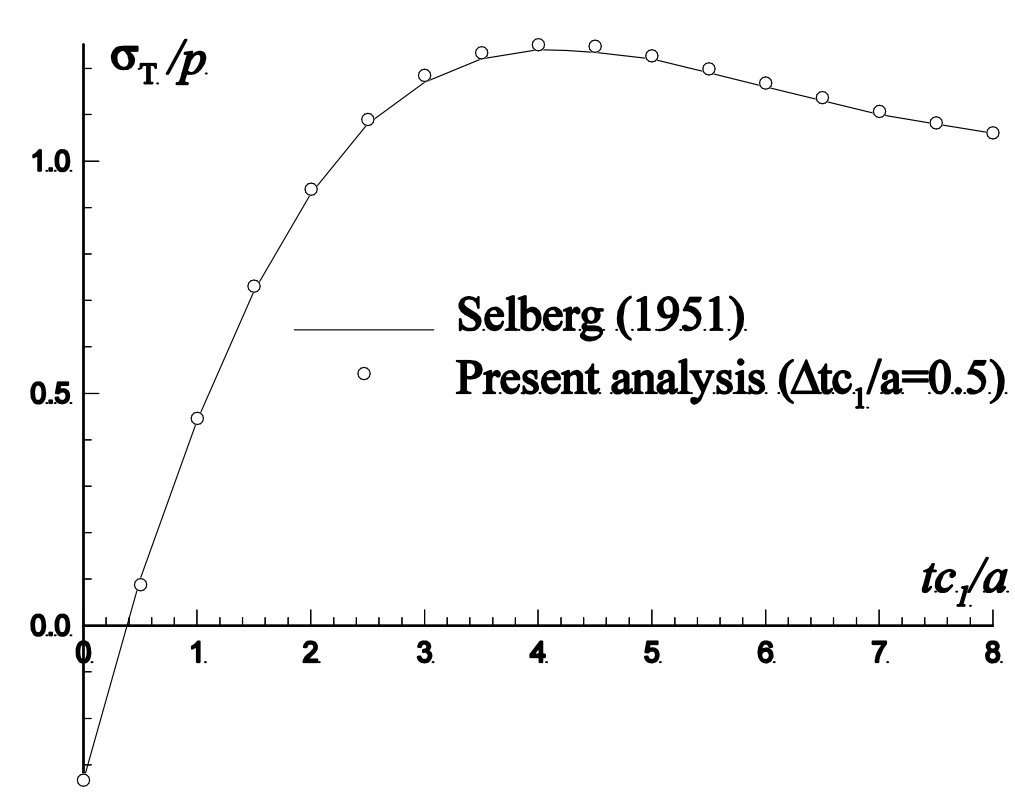

Figure 2: Normal stress

## Acknowledgement

Financial support from the University of Ferrara (FAR) is gratefully acknowledged.